%% file: main.tex
\PassOptionsToPackage{table}{xcolor} 
\documentclass{article} 
\usepackage[preprint]{template/colm2026_conference}

\usepackage{graphicx}       
\usepackage{microtype}      
\usepackage{hyperref}       
\usepackage{url}            
\usepackage{amsmath}        
\usepackage{amssymb}        
\usepackage{enumitem}       
\usepackage{xspace}         
\usepackage{arydshln}

\usepackage{booktabs}       
\usepackage{array}          
\usepackage{tabularx}       
\usepackage{longtable}      
\usepackage{siunitx}        
\usepackage{subfig}         

\usepackage{tcolorbox}      
\usepackage{xcolor}  
\usepackage{multirow}       
\usepackage{wrapfig}        
\usepackage{lineno}         
\usepackage{lipsum}         
\usepackage{nicematrix}     
\usepackage{adjustbox}
\usepackage{soul}
\usepackage{xcolor}

\definecolor{darkblue}{rgb}{0, 0, 0.5}
\hypersetup{colorlinks=true, citecolor=darkblue, linkcolor=darkblue, urlcolor=darkblue}

\usepackage{listings}

\definecolor{dark-gray}{gray}{0.85}
\definecolor{light-gray}{gray}{0.95}
\definecolor{mygreen}{rgb}{0,0.4,0}
\definecolor{mygray}{rgb}{0.5,0.5,0.5}
\definecolor{mymauve}{rgb}{0.58,0,0.82}
\definecolor{myred}{rgb}{0.82, 0.1, 0.26}

\usepackage[T1]{fontenc}
\usepackage{textcomp} 
\usepackage{xcolor}

\lstdefinestyle{CustomPy}{
    language=Python,
    upquote=true,               
    showstringspaces=false,
    basicstyle=\small\ttfamily,
    keywordstyle=\bfseries\color{green!40!black},
    commentstyle=\itshape\color{purple},
    stringstyle=\color{blue},   
    numbers=left,
    numberstyle=\tiny\color{gray},
    breaklines=true,
    columns=flexible,           
    escapeinside={(*@}{@*)}     
}

\title{From \swezero to \swehero: Execution-free to Execution-based Fine-tuning for Software Engineering Agents}

\author{Nikolai Ludwig\thanks{~Equal Contribution}, Wasi Uddin Ahmad\footnotemark[1], Somshubra Majumdar, Boris Ginsburg
\\ [1pt]
NVIDIA \\ [1pt]
Santa Clara, CA 95051, USA \\ [1pt]
\texttt{\{nliudvig, wasiuddina\}@nvidia.com} \\ [1pt]
\texttt{\url{https://huggingface.co/collections/nvidia/swe-zero-to-swe-hero}} 
}

\newcommand{\swezero}{\textsc{SWE-Zero}\xspace}
\newcommand{\swehero}{\textsc{SWE-Hero}\xspace}

\begin{document}

\setlength{\abovedisplayskip}{2pt}
\setlength{\belowdisplayskip}{2pt}
\setlength{\parskip}{2pt}

\ifcolmsubmission
\linenumbers
\fi

\maketitle

\begin{abstract}

We introduce \swezero to \swehero, a two-stage SFT recipe that achieves state-of-the-art results on SWE-bench by distilling open-weight frontier LLMs. Our pipeline replaces resource-heavy dependencies with an evolutionary refinement strategy: (1) \textsc{swe-zero} utilizes large-scale, execution-free trajectories to master code semantics and repository-level reasoning, and (2) \textsc{swe-hero} applies targeted, execution-backed refinement to transition these semantic intuitions into rigorous engineering workflows.

Our empirical results set a new benchmark for open-source models of comparable size. We release a dataset of 300k \swezero and 13k \swehero trajectories distilled from Qwen3-Coder-480B, alongside a suite of agents based on the Qwen2.5-Coder series. Notably, \textsc{swe-hero}-32B achieves a 62.2\% resolution rate on SWE-bench Verified. Furthermore, despite being trained exclusively on Python, our agents demonstrate robust zero-shot transferability on SWE-bench Multilingual, reaching 44.1\% and confirming the paradigm's generalizability across diverse languages.

\end{abstract}

\begin{figure*}[ht!]
\centering
\vspace{-3mm}
\includegraphics[width=0.95\textwidth]{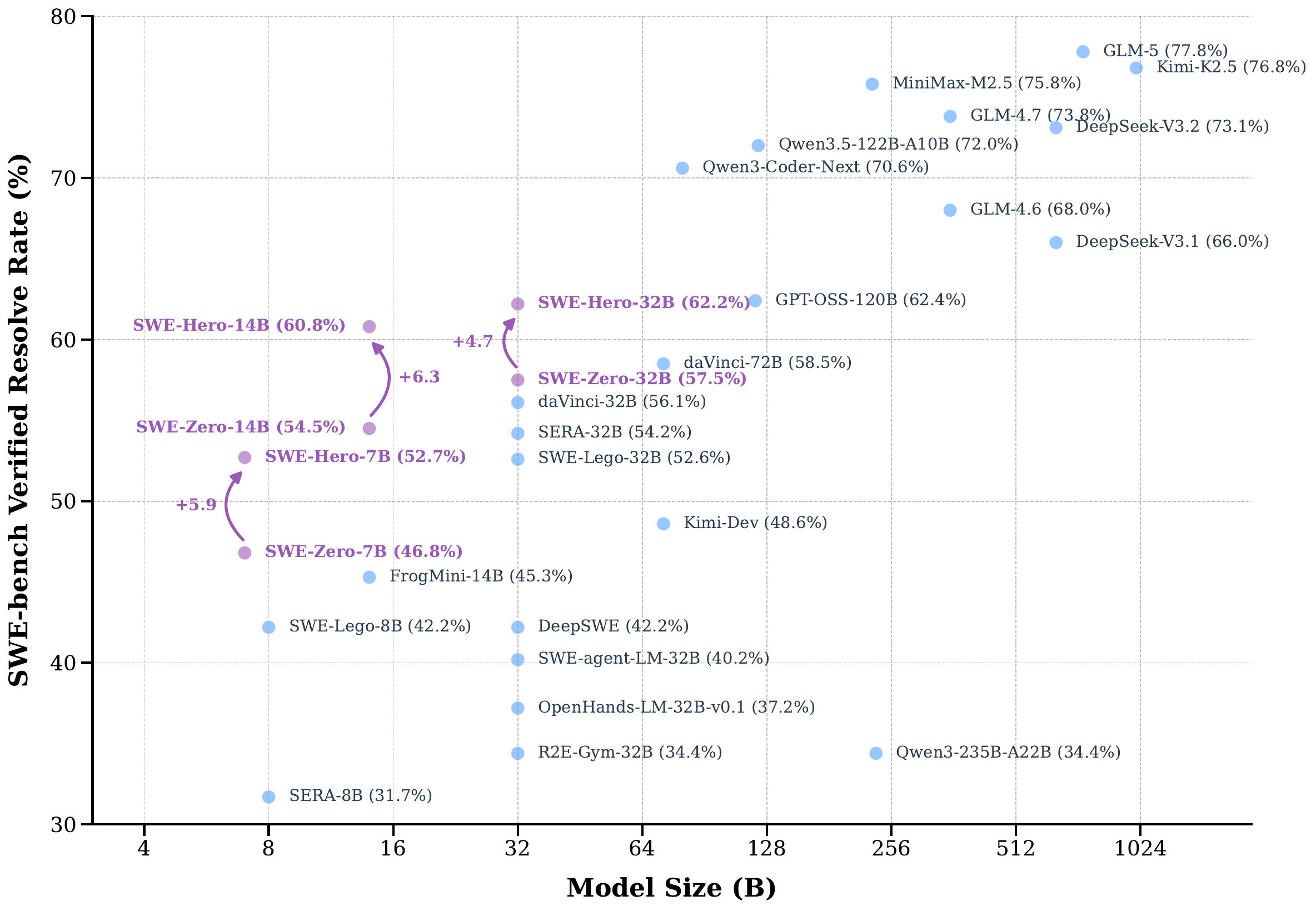}
\vspace{-2mm}
\caption{
Performance overview of various open-source foundational models and SWE agents on SWE-bench Verified. Our models, \swehero agents establish a new frontier, outperforming same-scale competitors.
}
\label{plot:result_highlight}
\vspace{-3mm}
\end{figure*}

\section{Introduction}
\label{sec:introduction}

\input{sections/1_introduction}

\section{Beyond Execution: Addressing the Infrastructure Bottleneck}
\label{sec:motivation}

\input{sections/2_motivation}

\section{Training Recipe for \swezero to \swehero}
\label{sec:method}

\input{sections/3_method}

\section{Experiments}
\label{sec:results}
\input{sections/4_evaluation}

\section{Ablation and Analyses}
\label{sec:ablation_and_analysis}

\input{sections/5_analysis}

\section{Related Works}
\label{sec:related_works}
\input{sections/6_related_works}

\section{Conclusion}
\label{sec:conclusion}

\input{sections/7_conclusion}

\clearpage



\section*{Limitations}

\paragraph{Model Inheritance and Bias} While our distillation recipe achieves state-of-the-art performance, it is subject to several inherent limitations. First, the \swezero and \swehero agents are distilled from Qwen3-Coder-480B; consequently, any underlying biases, stylistic preferences, or systemic errors present in the teacher model are likely inherited by our student models. Despite the performance gains observed, the student's ceiling is partially defined by the architectural and data-driven priors of its predecessor.

\paragraph{Architectural Scope} Furthermore, our current exploration focuses on standard (non-reasoning) LLM architectures. We recognize that the most recent high-performing models are increasingly reasoning-based, utilizing internal "Chain-of-Thought" or "thinking" processes to solve complex problems. While our current study does not evaluate these specialized reasoning models, we believe the \swezero-to-\swehero paradigm holds significant untapped potential for this new class of architectures. Future work could extend our two-stage distillation approach to reasoning-based LLMs, potentially unlocking even higher levels of autonomous problem-solving and systematic verification.

\paragraph{Stochastic and Environmental Variance} To ensure statistical reliability despite LLM stochasticity and the variability of complex execution environments, we conducted three independent training runs per model and report aggregate performance. For Test-Time Scaling (TTS), we utilized 32 rollouts per model to characterize the performance distribution. However, the intricacies of underlying infrastructure and dependency resolution in software environments introduce a degree of variance; consequently, specific scores may fluctuate across iterations, potentially impacting exact reproducibility.

\section*{Ethics Statement}
\paragraph{Dataset Integrity and Privacy}
We have implemented rigorous protocols to ensure the integrity and ethical standing of the \swezero and \swehero datasets, which will be released to the research community upon publication. All agent trajectories were distilled from open-source repositories governed by permissive licenses (e.g., MIT, Apache 2.0, or BSD), ensuring full compliance with original distribution rights. To protect individual privacy and maintain security, we employed automated filtering pipelines to identify and redact Personally Identifiable Information (PII), including developer emails, private access tokens, and sensitive credentials. By scrubbing these datasets of sensitive metadata prior to release, we aim to provide a high-utility resource that prioritizes data privacy and adheres to the ethical standards of open-source collaboration.

\paragraph{Declaration of Generative AI in the Writing Process}
During the preparation of this manuscript, the authors utilized Gemini 3 Flash to refine linguistic clarity and improve structural flow. The authors subsequently reviewed and edited all AI-assisted suggestions to ensure accuracy. The authors maintain full responsibility for the final content; the AI was utilized solely for stylistic enhancement and was not involved in data collection, analysis, or the formulation of original research findings.

\bibliography{bib/anthology,bib/colm2026_conference}
\bibliographystyle{template/colm2026_conference}

\input{sections/appendix}

\end{document}

%% file: sections/1_introduction.tex
The rapid evolution of large language model (LLM) agents has transitioned autonomous software engineering (SWE) from theoretical exploration to practical deployment \citep{yang2024swe, team2025kimi, cao2026qwen3}. Central to their success is an iterative interaction loop, where agents refine their outputs based on feedback from physical execution environments \citep{pan2024training, jain2025r2e}. In practice, this requires the instantiation of isolated, dependency-complete workspaces — typically via containerization frameworks like Docker — to ensure that code and unit tests execute reliably within the target repository's context. While these environments provide the verifiable runtime data necessary to capture high-quality trajectories from \emph{teacher} LLMs, this dependency on flawless execution has become a highly restrictive prerequisite for optimizing specialized SWE agents.

Despite their efficacy, current methodologies are constrained by a prohibitive computational overhead. Unlike algorithmic programming, which requires only a basic interpreter, real-world SWE tasks demand complex orchestration and brittle dependency resolution to verify even minor patches. To date, frontier research has relied almost exclusively on Docker-based paradigms for supervised fine-tuning (SFT) and reinforcement learning (RL). We argue that this absolute dependency on physical execution creates a multifaceted scalability bottleneck that constrains the development of open-source agents across three critical regimes:
\begin{itemize}[leftmargin=*, itemsep=0pt, topsep=2pt]
    \item {\bf Data Scalability}: A significant segment of real-world repositories and pull requests (PRs) are currently discarded because intricate or legacy configurations prevent them from building reliably within modern containerized environments.
    \item {\bf Training Scalability}: Orchestrating thousands of task-specific Docker images creates massive infrastructure overhead, complicating large-scale optimization and RL.
    \item {\bf Inference Scalability}: Expensive code execution and environment resets hinder agents from efficiently exploring solution branches using inference-time compute.
\end{itemize}

\input{tables/swe_zero_hero_intro}

To overcome these barriers, we propose a bifurcated training paradigm that decouples semantic reasoning from physical verification. 
We first introduce \swezero, an execution-free approach that leverages the internal "world models" of frontier LLMs to resolve issues without runtime feedback.
As demonstrated in Table \ref{tab:swe_zero_hero_intro_results}, frontier LLMs are now capable of resolving a significant percentage of SWE-bench Verified \citep{jimenez2023swe} issues — {\bf up to 69.5\%} — without any runtime feedback.
By eliminating any reliance on task-specific environments, \swezero unlocks the ``long tail'' of GitHub data previously inaccessible to execution-bound frameworks. 
We then bridge the gap via \swehero, a stage of grounded refinement through code execution.
By sequencing these stages, we transition the agent from an intuitive, execution-free understanding of repository architecture to a rigorous, execution-backed workflow. Our results demonstrate that agents resolve issues most effectively by internalizing code semantics before incurring the overhead of physical execution.

To this end, we propose {\bf \swezero to \swehero}, an \emph{efficient} SFT framework to train full-fledged SWE agents by first exposing them to large-scale, execution-free data, followed by a smaller, high-quality execution-based trajectory dataset. The framework employs a progressive data curriculum designed to transition an agent's learning from broad semantic reasoning to grounded, verifiable engineering:
\begin{enumerate}[leftmargin=*, itemsep=0pt, topsep=2pt]
    \item The \swezero Stage (Foundational Reasoning): We introduce the \swezero dataset, a collection of 300k agent trajectories generated {\bf without repository-specific code execution}. This dataset is constructed from 150k real-world GitHub pull requests (PRs) using a frontier open-source LLM \citep{yang2025qwen3} as an expert generator within the OpenHands scaffold \citep{wang2024openhands}. By employing strategic prompting and rigorous trajectory filtering, we utilize this data to train Qwen2.5-Coder models \citep{hui2024qwen2}. The resulting \swezero-Agent demonstrates competitive performance while consuming nearly 40\% fewer tokens by bypassing execution feedback cycles.

    \item The \swehero Stage (Grounded Refinement): We introduce the \swehero dataset, a refined collection of 13k trajectories across 13k task instances, supported by full containerized environments \citep{pan2024training, jain2025r2e, xie2025swe, badertdinov2025swe}. By fine-tuning the \swezero-Agent on this high-fidelity data, we produce models that outperform state-of-the-art models, particularly at the 7B and 14B parameter scales. To validate the generalization of our framework, we evaluate on SWE-bench Multilingual, where our \swehero-32B model achieves resolve rates of 44.1\% and 42.2\% (with and without execution, respectively).
\end{enumerate}

Furthermore, we provide a comprehensive analysis of the framework’s utility, including evaluations under test-time scaling (TTS), ablations on training sample density, and an efficiency audit of assistant turns and token consumption. We believe the \swezero to \swehero paradigm offers a scalable blueprint for building effective SWE agents. To support the research community, all associated data and models will be open-sourced.

%% file: tables/swe_zero_hero_intro.tex

\begin{table}[t]
\centering
\begin{tabularx}{0.99\linewidth}{X c cc}
\toprule
\textbf{Model} & \textbf{Execution} & \textbf{SWE-bench (V)} & \textbf{SWE-bench (M)} \\
\midrule
\multirow{2}{*}{MiniMax-M2.5}                    & $\times$      & 69.5 & 57.2 \\
                                                 & $\checkmark$  & 80.2 & 74.1 \\

\cdashline{1-4}[0.5pt/2pt] \noalign{\smallskip}
\multirow{2}{*}{Qwen3-Coder-Next}                & $\times$      & 56.9 & 50.7 \\
                                                 & $\checkmark$  & 71.3 & 64.3 \\
                                                 
\cdashline{1-4}[0.5pt/2pt] \noalign{\smallskip}                                              
\multirow{2}{*}{Qwen3-Coder-480B-A35B-Instruct}  & $\times$      & 59.4 & 44.3 \\
                                                 & $\checkmark$  & 69.6 & 54.7 \\

\cdashline{1-4}[0.5pt/2pt] \noalign{\smallskip}
\multirow{2}{*}{\swehero-32B (Ours)}             & $\times$      & 57.7 & 42.2 \\
                                                 & $\checkmark$  & 62.2 & 44.1 \\
\bottomrule
\end{tabularx}
\vspace{-2mm}
\caption{
Resolve rates (\%) on SWE-bench Verified (V) and Multilingual (M). We compare the execution-free ($\times$) paradigm against the execution-backed setup ($\checkmark$) using OpenHands. Notably, our \swehero-32B model is fine-tuned in two successive SFT stages using a collection of 300k execution-free and 13k execution-based Python trajectories.
}
\vspace{-2mm}
\label{tab:swe_zero_hero_intro_results}
\end{table}

%% file: sections/2_motivation.tex
The resolution of software engineering (SWE) issues is traditionally viewed as an interactive, execution-heavy process. In this paradigm, an agent must navigate a codebase, identify a bug, and iteratively verify potential fixes through a feedback loop with a live environment. However, we argue that this heavy reliance on physical execution creates a massive scalability bottleneck that hinders the development of open-source agents.

\paragraph{The Four Layers of SWE Environments}
To quantify the operational complexity inherent to SWE-agents, we deconstruct the SWE environment into four hierarchical layers, each increasing in utility at the cost of maintenance and compute:
\begin{itemize}[leftmargin=*, itemsep=0pt, topsep=2pt]
    \item File System: The foundational layer comprising the repository’s raw files and directories; the primary state an agent must navigate and modify to resolve an issue.
    
    \item Terminal: This layer provides the basic command-line interfaces (e.g., \verb|ls|, \verb|grep|, \verb|cat|) required to navigate and inspect the file system.
    
    \item Sandbox: A controlled environment that couples the file system with a terminal to support deterministic navigation and editing. It supports "lightweight" exploration but lacks the dependencies to compile or run the software.
    
    \item Docker (Task-Specific Environment): The most resource-intensive layer, which instantiates a dependency-complete container for physical verification. While it allows for running unit tests, its maintenance is notoriously brittle and imposes massive computational and storage overhead during large-scale training.
\end{itemize}

\paragraph{Scaling via \swezero: Execution-Free Resolution}
Existing approaches depend heavily on the task-specific environment layer for optimization. In practice, this forces the discard of many real-world repositories because their complex configurations prevent them from building reliably in a container. Our framework, \swezero, introduces a paradigm shift toward execution-free synthesis—an approach to software engineering that operates independently of a runtime environment. This method leverages the observation that frontier LLMs have developed a sophisticated internal "world model" of code semantics, allowing them to map complex repository architectures and resolve issues without external feedback. 
To avoid the prohibitive overhead of verifiable environments, \swezero agents internally simulate the software engineering workflow through:
\begin{itemize}[leftmargin=*, itemsep=0pt, topsep=2pt]
    \item Direct Bug Identification: Pinpointing root causes through contextual reasoning without needing to reproduce failures through physical execution.
    
    \item Semantic-Driven Patching: Generating fixes based on an internalized understanding of codebase architecture, bypassing the resource-heavy trial-and-error cycle of unit tests.
    
    \item Infrastructure-Agnostic Scaling: Leveraging a massive volume of distilled trajectories, unrestricted by the need for containerized build environments.
\end{itemize}

\paragraph{Bridging the Gap: From Intuition to Verification}
While \swezero enables massive scaling, an execution-free approach lacks the safety net of physical feedback. This "gut feeling" for software patterns can struggle with non-obvious bugs, such as race conditions, that require observing runtime behavior. We argue the most effective frontier agent is built by \textbf{sequencing} these paradigms rather than choosing between them:
\begin{itemize}[leftmargin=*, itemsep=0pt, topsep=2pt]
    \item \swezero (Foundational SFT): The agent learns repository-level reasoning at scale through a zero-infrastructure curriculum.
    \item \swehero (Grounded Refinement): A targeted, execution-backed dataset transitions the model from conceptual intuition to a rigorous, verified engineering workflow.
\end{itemize}

%% file: sections/3_method.tex
We build upon the SWE-agent paradigm \citep{yang2024swe}, in which LLMs operate via an interface that enables autonomous repository navigation, code editing, and test execution. Within this framework, we define a Task Instance as a specific codebase issue consisting of a natural-language problem description, a repository with a base commit to use as the agent's starting point, associated test cases, and a reference "golden" patch. Complementary to this, a Trajectory represents the chronological sequence of an agent's actions interleaved with observations from the environment. Our primary objective is to curate a dual-tier training resource by developing a massive, execution-free corpus to provide broad semantic coverage, which is then supplemented by a high-fidelity, small-scale dataset of trajectories grounded in physical, execution-based feedback.

\paragraph{Task and Repository Collection.} We aggregate several open-source Python SWE datasets \citep{pan2024training, jain2025r2e, badertdinov2025swe, xie2025swe}, consolidating over 180k task instances across $3,500$+ repositories with permissive licenses. From this collection, 13.5k instances include containerized Docker environments provided by the original authors, which we verify by executing their reference patches. These two distinct tiers—the vast, unconstrained task instances and the verified, containerized subset—serve as the foundation for the \swezero and \swehero datasets, respectively.

\paragraph{Agent Scaffolding.} We use OpenHands \citep{wang2024openhands}, an open-source platform as the agent framework. OpenHands enables agents to iteratively edit files and execute shell commands—a setup proven to yield robust, reproducible results on benchmarks like SWE-bench. For our experiments, we equip the agent with four tools: \texttt{str\_replace\_editor} for file reading and editing, \texttt{execute\_bash} for command execution, \texttt{think} for longer reasoning between actions, and \texttt{finish} to signal task completion and submit the solution.

\paragraph{Teacher Agent.} We employ Qwen3-Coder-480B-A35B-Instruct \citep{yang2025qwen3} as our teacher model and enforce strict security constraints to ensure evaluation integrity. This open-weight LLM provides state-of-the-art coding capabilities suitable for large-scale trajectory distillation. To prevent "git hacking" and data leakage \citep{xiao2026mimo}, we remove all Git commits, tags and branches created after the base commit for each task instance, ensuring the agent cannot retrieve ground-truth solutions from future repository states.

\input{prompts/swezero_vs_swehero}

\subsection{ Trajectory Synthesis and Data Curation}

This section details the trajectory generation and filtering pipeline for our framework. The structural difference between the \swezero and \swehero trajectory generation processes is illustrated in Fig. \ref{fig:vertical_setup_comparison}. We prompt the teacher agent across two distinct configurations: one provided with a full execution environment for the given task and another restricted to a generic sandbox with no repository-specific setup. To compensate for the lack of runtime feedback in the latter, we implement a multi-stage filtering pipeline to ensure all generated trajectories remain coherent and logically sound.

\paragraph{\swezero Trajectory Collection}

In this configuration, the teacher agent is provided with a problem statement, the corresponding repository checked out to the task's base commit, and a sandbox. Critically, the sandbox image is identical across all task instances and lacks repository-specific environment setups. Deprived of a task-specific environment, the agent is restricted to codebase exploration and source file modification without the capacity for code execution.
These trajectories typically progress through five distinct phases: {\bf requirements analysis, repository exploration, fix localization, patch implementation, and final review}. Because the agent cannot rely on iterative runtime feedback, the resulting trajectories are significantly more concise; empirically, these execution-free trajectories contain 40\% fewer tokens than their execution-backed counterparts.
We generate $N$ rollouts ($N \in [3, 5]$) per task instance, establishing a diverse foundational pool for subsequent filtering and distillation.

\paragraph{Trajectory Filtering} While the teacher LLM is highly effective in execution-enabled environments, our curriculum requires it to operate within a zero-execution constraint. In this setting, the model may occasionally disregard system instructions and attempt to invoke testing or execution tools. Because the \swezero setup lacks a task-specific environment, these attempts receive no productive runtime feedback, which can lead to infinite loops or non-deterministic behaviors that compromise rollout quality. To mitigate this, we employ a multi-stage filtering pipeline. In the first stage, we use a rule-based parser (see Fig. \ref{fig:bash_parser}) to inspect all tool calls and automatically discard any trajectories where the LLM attempted prohibited code execution. Applying these criteria pruned 35\% of the initial rollouts, ensuring the remaining corpus strictly adheres to execution-free constraints.

In a second filtering stage, we apply a rigorous quality-control mechanism to ensure the integrity of the training corpus. The pipeline automatically discards trajectories that exceed the maximum step limit or result in null code changes. To prevent ``shortcut'' behaviors, we filter out trajectories where the agent modifies any file present in the test patch. Finally, to encourage correct tool call usage, we prune trajectories where the agent fails to produce exactly one tool call per turn or commits three or more errors when invoking the \verb|str_replace_editor| tool.

\paragraph{\swehero Trajectory Collection}
In this configuration, the teacher agent operates within a full Docker environment, performing a standard trajectory to resolve software issues with real-time feedback. Unlike the \swezero rollouts, which require tool-use validation, \swehero trajectories proceed directly to the second filtering stage to ensure high-fidelity reasoning and successful issue resolution. We generate a single rollout per task instance, resulting in a concentrated candidate pool for final model refinement.

\paragraph{Trajectory Distribution and Corpus Composition}
In the \swehero configuration, trajectory lengths are predominantly concentrated between 20 and 60 interaction turns; conversely, \swezero trajectories are significantly more concise, typically spanning 10 to 30 turns. Our initial Stage-1 generation produced approximately 1M+ trajectories, which was reduced to 600k following the filtering process. For our final training corpus, we selected two trajectories per task, resulting in a collection of 300k \swezero trajectories across 150k task instances. We found that the remaining 30k instances did not yield any trajectories that met our rigorous filtering criteria. In contrast, the high-fidelity \swehero set comprises 13.2k execution-based trajectories following identical filtering protocols. Notably, due to the smaller scale of the \swehero dataset, we do not exclude trajectories based on whether the associated tasks were successfully resolved. 

\subsection{Supervised Fine-Tuning (SFT)} 

We perform multi-turn SFT on Qwen2.5-Coder-Instruct models using the filtered dataset and corresponding trajectories. To support long-form multi-turn modeling, we apply YaRN \citep{peng2023yarn} to extend the base context length from 32k to 128k tokens. Our training process employs a multi-turn masking strategy that excludes tool outputs from the loss computation, ensuring the model prioritizes action generation over fitting execution outputs. Training on the \swezero instances produces the base LLM agents, which then serve as the initialization for the subsequent \swehero training stage.

%% file: prompts/swezero_vs_swehero.tex

\begin{figure*}[t!]
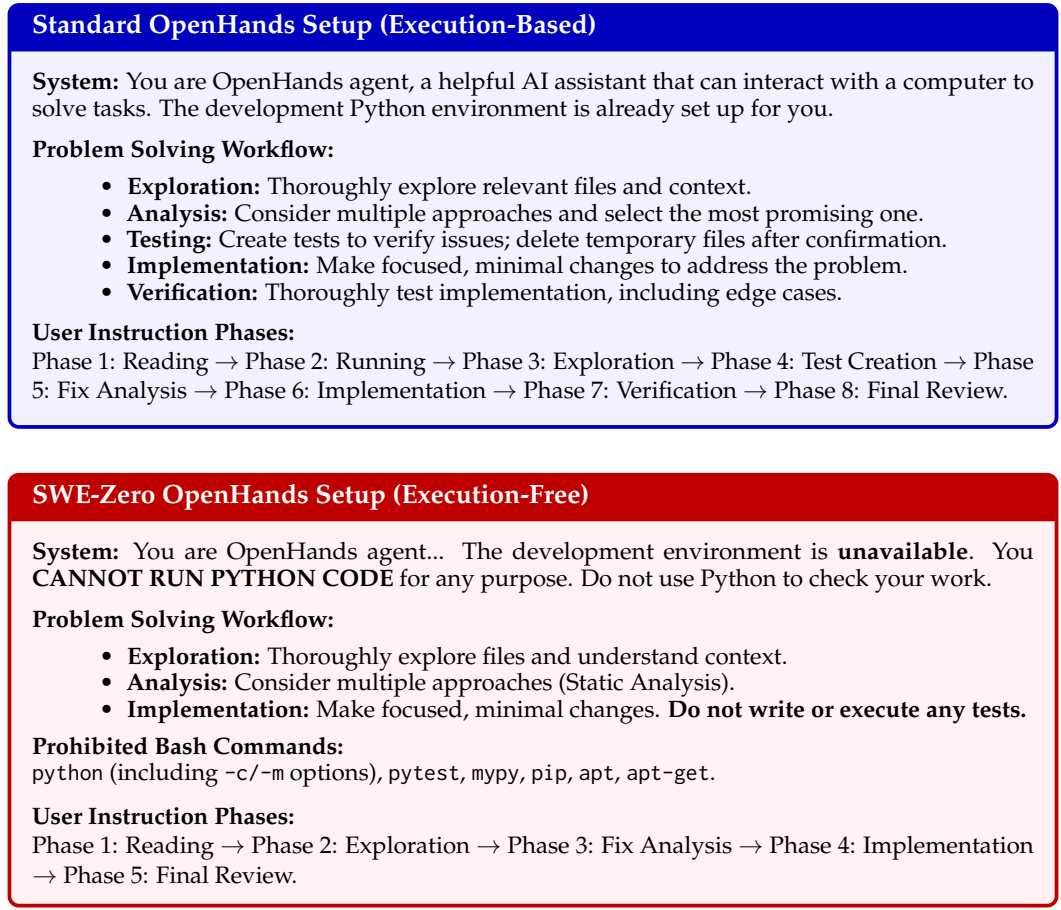

\centering

\begin{tcolorbox}[title={Standard OpenHands Setup (Execution-Based)}, 
    colback=blue!5, colframe=blue!75!black, fonttitle=\bfseries,
    left=5pt, right=5pt, top=5pt, bottom=5pt]
{\small
\textbf{System:} You are OpenHands agent, a helpful AI assistant that can interact with a computer to solve tasks. The development Python environment is already set up for you.

\vspace{0.2cm}
\textbf{Problem Solving Workflow:}
\begin{itemize}[noitemsep]
    \item \textbf{Exploration:} Thoroughly explore relevant files and context.
    \item \textbf{Analysis:} Consider multiple approaches and select the most promising one.
    \item \textbf{Testing:} Create tests to verify issues; delete temporary files after confirmation.
    \item \textbf{Implementation:} Make focused, minimal changes to address the problem.
    \item \textbf{Verification:} Thoroughly test implementation, including edge cases.
\end{itemize}

\textbf{User Instruction Phases:} \\
Phase 1: Reading $\rightarrow$ Phase 2: Running $\rightarrow$ Phase 3: Exploration $\rightarrow$ Phase 4: Test Creation $\rightarrow$ Phase 5: Fix Analysis $\rightarrow$ Phase 6: Implementation $\rightarrow$ Phase 7: Verification $\rightarrow$ Phase 8: Final Review.
}
\end{tcolorbox}

\vspace{0.2cm} 

\begin{tcolorbox}[title={SWE-Zero OpenHands Setup (Execution-Free)}, 
    colback=red!5, colframe=red!75!black, fonttitle=\bfseries,
    left=5pt, right=5pt, top=5pt, bottom=5pt]
{\small
\textbf{System:} You are OpenHands agent... The development environment is \textbf{unavailable}. You \textbf{CANNOT RUN PYTHON CODE} for any purpose. Do not use Python to check your work.

\vspace{0.2cm}
\textbf{Problem Solving Workflow:}
\begin{itemize}[noitemsep]
    \item \textbf{Exploration:} Thoroughly explore files and understand context.
    \item \textbf{Analysis:} Consider multiple approaches (Static Analysis).
    \item \textbf{Implementation:} Make focused, minimal changes. \textbf{Do not write or execute any tests.}
\end{itemize}

\textbf{Prohibited Bash Commands:} \\
\texttt{python} (including \texttt{-c/-m} options), \texttt{pytest}, \texttt{mypy}, \texttt{pip}, \texttt{apt}, \texttt{apt-get}.

\vspace{0.2cm} 
\textbf{User Instruction Phases:} \\
Phase 1: Reading $\rightarrow$ Phase 2: Exploration $\rightarrow$ Phase 3: Fix Analysis $\rightarrow$ Phase 4: Implementation $\rightarrow$ Phase 5: Final Review.
}
\end{tcolorbox}
\vspace{-2mm}
\caption{Comparative overview of Agentic Workflows in OpenHands. The bottom panel depicts the SWE-Zero setup, utilizing a restricted, execution-free environment to maximize data scalability. In contrast, the top panel illustrates the SWE-Hero configuration, which transitions to a standard, feedback-driven workflow grounded in physical execution.}
\vspace{-2mm}
\label{fig:vertical_setup_comparison}
\end{figure*}

%% file: sections/4_evaluation.tex
\input{tables/main}

\subsection{Experiment Setup}

\paragraph{Evaluation Benchmarks and Metrics.}
We evaluate our models on two primary benchmarks using Resolved Rate (\%) as the core metric: (1) SWE-bench Verified  \citep{jimenez2023swe}: A human-curated subset of 500 high-quality, real-world Python software engineering issues; and (2) SWE-bench Multilingual\footnote{\url{https://huggingface.co/datasets/SWE-bench/SWE-bench_Multilingual}}: A 300-task benchmark focused on rapid evaluation of cross-language generalization. All reported results represent the average of three evaluation passes to account for variance in model outputs.

\paragraph{Model Training and Inference Setup.} 
We developed \swezero and \swehero in 7B, 14B, and 32B variants by fine-tuning Qwen2.5-Coder-Instruct for up to three epochs, using a maximum context length of 128k tokens and a global batch size of 32.  We employ a cosine learning rate scheduler that decayed from a peak of 1e-5 to 1e-8 after a 0.1 warmup ratio. For final evaluation, we fixed the inference temperature at 0.7, top-p at 0.8, and top-k at 20. Furthermore, inference was performed with a maximum context capacity of 128k tokens and a limit of 100 interaction rounds per task.

\subsection{Experiment Results}

Table \ref{tab:swe-bench-results} and \ref{tab:swe_zero_hero_intro_results} shows the results of our models and the baselines on SWE-bench Verified and Multilingual benchmarks. We can obtain the following observations:

\begin{itemize}[leftmargin=*, itemsep=2pt, topsep=2pt]
    \item \emph{Redefining the Open-Source Frontier on SWE-bench.} The \swehero agents demonstrate superior performance relative to existing open-source code agents across multiple scales. Our 7B, 14B, and 32B models achieve resolution rates of 52.7\%, 60.8\%, and 62.2\%, respectively. Notably, the 7B and 14B variants outperform contemporary open-source agents of equivalent size by a significant margin, establishing a new efficiency frontier for smaller-scale models. Our 32B model delivers performance competitive with the current state-of-the-art, OpenSWE-32B \citep{fu2026davincienvopensweenvironment}, despite a substantially more streamlined training process. While OpenSWE-32B relies on a resource-intensive setup—incorporating 45,000 executable Docker environments across 12,800 repositories—SWE-Hero achieves comparable results through our more scalable distillation recipe, proving that high-tier agentic reasoning can be instilled without prohibitive infrastructure costs.

    \item \emph{Effectiveness of \swezero to \swehero Training.} To demonstrate the critical role of the \swezero stage, we conduct an ablation study by fine-tuning the Qwen2.5-Coder-32B-Instruct model directly on the \swehero dataset, bypassing the execution-free phase. \textbf{This direct-to-hero baseline achieves resolve rates of only 55.7\% on SWE-bench Verified and 30.8\% on SWE-bench Multilingual}. In contrast, incorporating the foundational \swezero stage boosts these figures to 62.2\% and 44.1\%, respectively. These results provide empirical proof that the \swezero stage is not only computationally efficient due to its lack of execution overhead, but also serves as a critical prerequisite that provides a strong inductive bias for the subsequent \swehero refinement.

    \item \emph{Unlocking Cross-language Generalization.} The impact of our training recipe is particularly pronounced in its capacity for cross-language transfer. As shown in Table \ref{tab:swe_zero_hero_intro_results} (full results in Table \ref{tab:swe-bench-multi-results-full} in Appendix), \swehero-32B demonstrates a significant performance leap on SWE-bench Multilingual compared to the direct-to-hero baseline. Despite being trained primarily on Python trajectories, the foundational reasoning patterns distilled during the \swezero phase enable the model to generalize effectively across diverse programming languages thanks to the large-scale \swezero stage. By distilling 300k execution-free trajectories, this phase provides a massive breadth of foundational reasoning patterns that enable robust cross-language transfer, proving that scale in the initial distillation phase is the primary driver of our framework's generalizability.

\end{itemize}

%% file: tables/main.tex
\begin{table}[ht!]
\centering
\resizebox{1.0\linewidth}{!} {%
\begin{tabular}{llcc}
\toprule
\textbf{Model} & \textbf{Scaffold} & \textbf{Training} & \textbf{Resolve Rate (\%)} \\ 



\midrule

\rowcolor{gray!10} \textit{Parameters $\approx$ 7B} & & & \\
SWE-Gym-7B \citep{pan2024training} & OpenHands & SFT & 10.6 \\
R2E-Gym-7B \citep{jain2025r2e} & R2E-Gym & SFT & 19.0 \\
SWE-Mirror-LM-7B \citep{wang2025swe} & MOpenHands & SFT & 22.8 \\
SWE-Dev-7B \citep{wang2025swe} & OpenHands & RL & 23.4 \\
SERA-8B \citep{shen2026sera} & SWE-agent & SFT & 31.7 \\
SWE-Lego-Qwen3-8B & OpenHands & SFT & 42.2$^\dagger$ \\
\rowcolor{cyan!5} SWE-Zero-7B (Ours) & OpenHands & SFT & {\bf 46.8}$^\dagger$ \\
\rowcolor{cyan!5} SWE-Hero-7B (Ours) & OpenHands & SFT & {\bf 52.7}$^\dagger$ \\
\hline

\rowcolor{gray!10} \textit{Parameters $=$ 14B} & & & \\
SWE-Gym-14B \citep{pan2024training}  & OpenHands & SFT & 16.4 \\
Qwen3-14B \citep{yang2025qwen3} & OpenHands & - & 17.3 \\
R2E-Gym-14B \citep{jain2025r2e} & R2E-Gym & SFT & 26.8 \\
FrogMini‑14B \citep{sonwane2025bugpilot} & R2E‑Gym & SFT+RL & 45.3 \\
\rowcolor{cyan!5} SWE-Zero-14B (Ours) & OpenHands & SFT & {\bf 54.5}$^\dagger$ \\
\rowcolor{cyan!5} SWE-Hero-14B (Ours) & OpenHands & SFT & {\bf 60.8}$^\dagger$ \\
\hline

\rowcolor{gray!10} \textit{Parameters $=$ 32B} & & & \\
SWE-Gym-32B \citep{pan2024training}  & OpenHands & SFT & 20.6 \\
R2E-Gym-32B \citep{jain2025r2e} & R2E-Gym & SFT & 34.4 \\
SWE-Dev-32B \citep{wang2025swe} & OpenHands & RL & 36.6 \\
Skywork-SWE-32B \citep{zeng2025skywork} & OpenHands & SFT & 38.0 \\
DeepSWE-32B-Preview \citep{luo2025deepswe} & OpenHands & RL & 42.2 \\
SWE-Mirror-LM-32B \citep{wang2025swe} & MOpenHands & SFT & 52.2 \\
SWE-Lego-Qwen3-32B \citep{tao2026swe} & OpenHands & SFT & 52.6$^\dagger$ \\
SERA-32B \citep{shen2026sera} & SWE-agent & SFT & 54.2 \\
FrogBoss‑32B \citep{sonwane2025bugpilot} & R2E‑Gym & SFT+RL & 54.6 \\
daVinci-Dev-32B \citep{zeng2026davinci} & SWE-Agent & SFT & 56.1 \\
\rowcolor{cyan!5} SWE-Zero-32B (Ours) & OpenHands & SFT & {\bf 57.5}$^\dagger$ \\
SWE-Swiss-32B \citep{he2025sweswiss} & Agentless & SFT+RL & 58.0 \\
SWE-Master-32B-RL \citep{song2026swe} & R2E-Gym & SFT+RL & 61.4 \\
\rowcolor{cyan!5} SWE-Hero-32B (Ours) & OpenHands & SFT & {\bf 62.2}$^\dagger$ \\
OpenSWE-32B \citep{fu2026davincienvopensweenvironment} & SWE-Agent & SFT & 62.4 \\
\bottomrule
\end{tabular}
}
\vspace{-2mm}
\caption{Performance comparison on the SWE-bench Verified. Results marked with $^\dagger$ exclude \emph{Git hacking}; for all other scores, the use of \emph{Git hacking} is unspecified.}
\vspace{-2mm}
\label{tab:swe-bench-results}
\end{table}

%% file: sections/5_analysis.tex
\paragraph{The Success-to-Cost Trade-off: Efficiency vs. Accuracy}

\begin{figure*}[ht!]
\centering
\vspace{-1mm}
\includegraphics[width=0.99\textwidth]{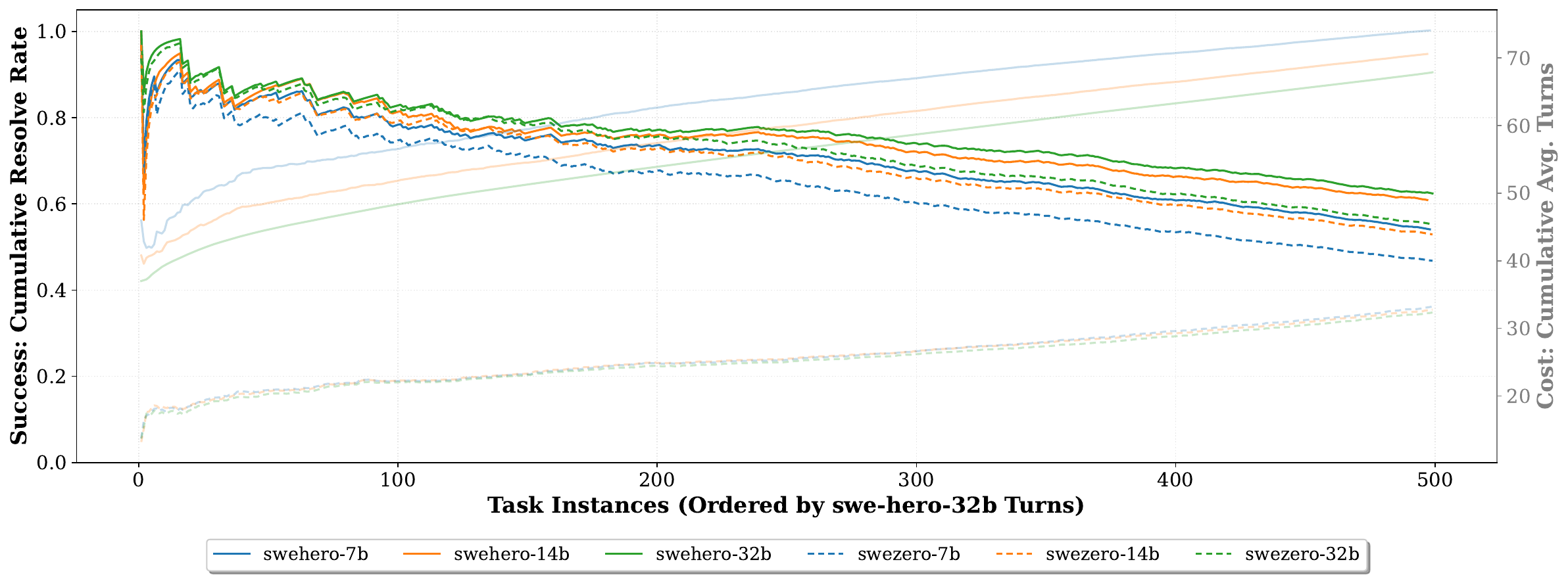}
\vspace{-1mm}
\caption{
Cumulative performance and resource efficiency on SWE-bench Verified. Tasks are ordered by \swehero-32b turn-count (complexity proxy). The primary axis (thick lines) shows the natural decline in resolve rate as complexity grows, while the secondary axis (faint lines) highlights the significantly lower turn-costs of \swezero models.
}
\label{plot:tasks_vs_turns}
\vspace{-2mm}
\end{figure*}

Figure~\ref{plot:tasks_vs_turns} illustrates a clear divergence in trajectory dynamics between the two frameworks. \swezero (Static Reasoning) models resolve tasks in significantly fewer turns by bypassing the execution and test-validation cycle. Operating as high-throughput, execution-free agents, they rely on distilled reasoning to follow a "direct path" to a solution. In contrast, \swehero adopts the standard iterative validation paradigm. By integrating execution-based SFT, these agents utilize additional turns to author and run tests, establishing a robust feedback loop. This systematic verification explains the higher turn counts; the agent prioritizes reliability over speed by "double-checking" patches against runtime results. Ultimately, \swehero achieves superior success rates on the most challenging tasks by trading temporal efficiency for a significant gain in resolution accuracy.

\paragraph{Test-time Scaling (TTS)}
We evaluate Test-time Scaling (TTS) by generating $K=32$ candidate rollouts and employing open-source generative verifiers \citep{pan2024training, jain2025r2e, tao2026swe} to select the optimal patch. While this approach yields significant resolution gains—up to 7.9\% for our smaller models and 4.5\% for the 32B variant—a substantial gap remains between the verifier-selected Best@K and the theoretical Pass@K ceiling. This margin, which reaches 15\% at $K=16$ and continues to widen with larger $K$, suggests that current open-source verifiers lack the discriminative precision necessary to fully exploit TTS. We identify this selection bottleneck as a critical challenge and leave the development of robust reward models for future work.
A comprehensive analysis of these scaling dynamics and individual verifier benchmarks is provided in Appendix \ref{subsec:appendix_tts}.

\paragraph{Data Scaling Laws in \swezero Training}

\begin{wrapfigure}{r}{0.5\textwidth}
  \centering
  \includegraphics[width=0.5\textwidth]{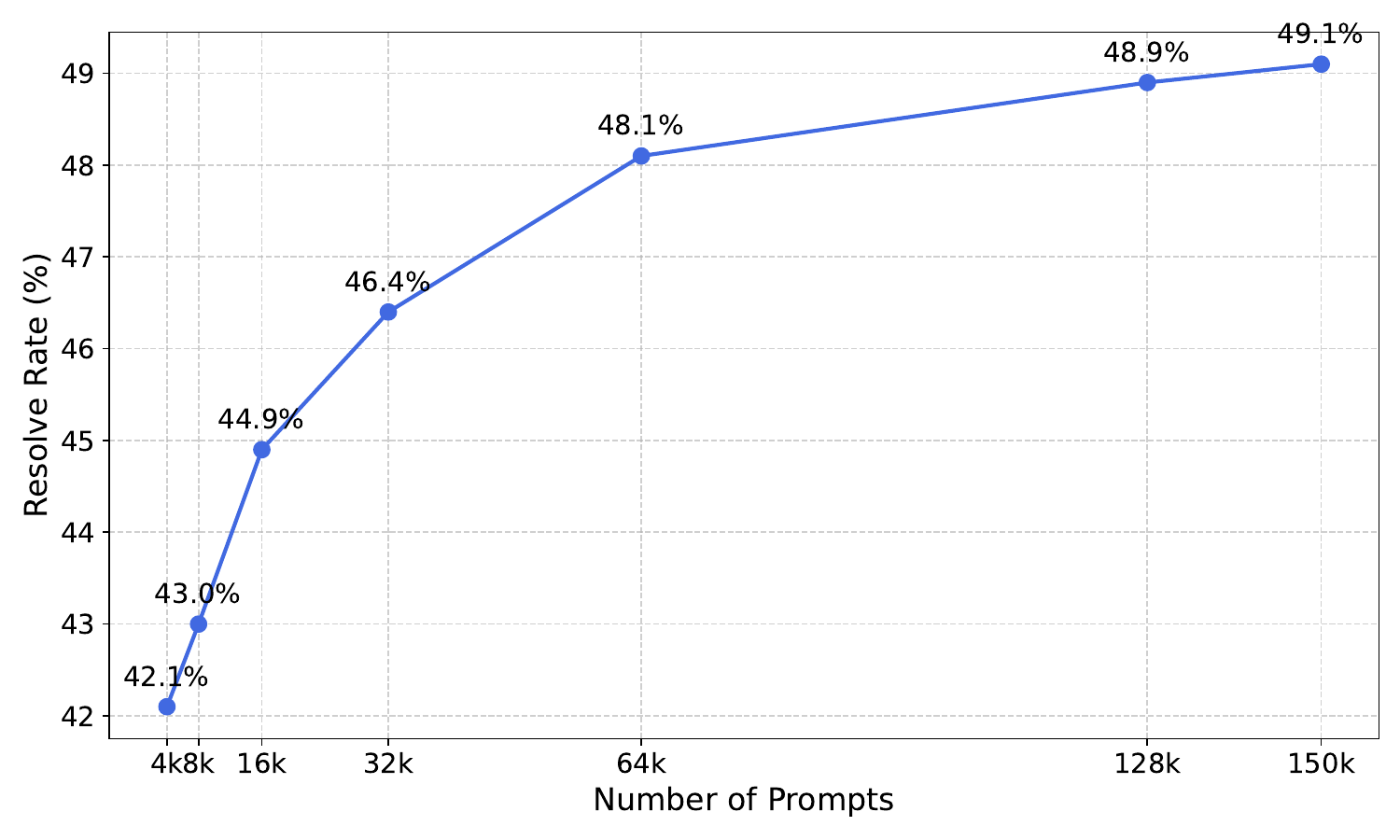}
  \vspace{-6mm}
  \caption{Scaling behavior of \swezero-14B. The resolution rate on SWE-bench increases consistently as execution-free training samples scale from 4k to 150k.}
  \label{fig:scaling_laws}
  \vspace{-2mm}
\end{wrapfigure}

As shown in Figure \ref{fig:scaling_laws}, our ablation of \swezero-14B highlights the impact of data volume on agent efficacy. Scaling from 4k to 150k GitHub PRs improves the resolution rate from 42.1\% to 49.1\%.
We observe the most rapid gains in the lower-data regime (4k to 32k), suggesting that the model quickly internalizes foundational understanding patterns early in the scaling process. While the performance curve exhibits diminishing returns at higher volumes, the consistent growth up to 150k samples provides clear empirical evidence that scaling distilled trajectories is a primary driver of model performance. This underscores the efficiency of \swezero, enabling significant bootstrapping through data volume without the overhead of grounded execution.

%% file: sections/6_related_works.tex

\paragraph{Software Issue Resolving Datasets.} 
The evolution of issue-resolving datasets has transitioned from static code collections to dynamic, interactive environments. SWE-Gym \citep{pan2024training} pioneered the pairing of Python issues with executable environments for agent training, a paradigm later refined by SWE-rebench \citep{badertdinov2025swe} and SWE-Factory \citep{guo2025swe} to combat data contamination through continuous sourcing. To address the manual curation bottleneck, SWE-Smith \citep{yang2025swe} and SWE-Synth \citep{pham2025swe} introduced automated bug injection and debugging simulation. This scalability is further enhanced by R2E-Gym \citep{jain2025r2e}, which curates environments from code commits, and SWE-Mirror \citep{wang2025swe}, which re-animates the semantic essence of real-world issues. Notably, SWE-World \citep{sun2026swe} departs from physical execution by using a learned surrogate model to predict environment feedback. Synthesizing these trends, SWE-Lego \citep{tao2026swe} merges high-quality real-world pull requests with scalable synthetic instances to optimize both the precision and volume of agent training.

\paragraph{Software Engineering Agents.} 
To enhance autonomous code agents, research has transitioned from framework development to model-centric optimization. Early systems like SWE-agent \citep{yang2024swe} and OpenHands \citep{wang2024openhands} established agentic frameworks using frontier models \citep{liu2025deepseek, cao2026qwen3} in grounded environments. Recent work, however, focuses on instilling agentic reasoning directly into base models via data synthesis and mid-training, as seen in daVinci-Dev \citep{zeng2026davinci}, or through agentic SFT using high-quality trajectories, as demonstrated by SWE-Mirror \citep{wang2025swe} and SWE-Lego \citep{tao2026swe}. To bridge the gap between static training and dynamic interaction, DeepSWE \citep{luo2025deepswe} and SkyRL \citep{cao2025skyrl} employ RL \citep{golubev2025training, song2025r1} to optimize policies through trial-and-error feedback. Parallel to these end-to-end autonomous loops, ``agentless'' pipelines \citep{xia2024agentless, yang2025kimi} decompose issue-solving into discrete, optimized stages: fault localization, code repair, and patch verification \citep{he2025sweswiss, xie2025swe}.

%% file: sections/7_conclusion.tex
In this work, we introduced a novel SFT recipe, transitioning from \swezero to \swehero, designed to overcome the scalability bottlenecks of traditional automated software issue resolution. By shifting away from a total reliance on labor-intensive, verifiable execution environments, we have demonstrated that a bifurcated distillation approach—combining large-scale execution-free fine-tuning with targeted execution-based refinement—can achieve state-of-the-art performance. To facilitate further advancement in the field, we open-source our comprehensive dataset and the Qwen2.5-Coder-based agent suite, providing a scalable framework for autonomous software resolution. 

%% file: sections/appendix.tex
\appendix
\clearpage
{
\centering
\Large\bf Supplementary Material: Appendices \\ [5pt]
}

\section{Additional Analyses}

\subsection{Resilience to Repository Metadata Leakage}
Recent studies \citep{tao2026swe, xiao2026mimo, song2026swe} have highlighted a "Git hacking" vulnerability where LLM agents exploit repository metadata—such as commit logs or historical diffs—to bypass reasoning and directly locate ground-truth patches. While current execution-based frameworks must implement complex sanitization measures to secure the environment, \swezero is inherently resilient to this class of leakage. Because the \swezero paradigm operates without a task-specific containerized environment or live Git history, the agent has no access to the underlying metadata required for such exploits. This design provides a significant structural advantage: it eliminates the "infrastructure pitfalls" associated with verifiable environments while ensuring that model performance is driven by genuine semantic reasoning rather than metadata shortcuts.

\subsection{Test-time Scaling (TTS)}
\label{subsec:appendix_tts}
Test-time scaling (TTS) enhances the performance of software agents by allocating additional computational resources during inference. This is generally achieved through two complementary dimensions: sequential scaling, which increases the number of interaction turns, and parallel scaling, which generates multiple independent rollouts and selects the optimal solution via a verifier. In this study, we focus on parallel scaling, utilizing a verifier to assign a score ($s \in [0, 1]$) to each trajectory and its proposed patch based on the likelihood of a successful resolution.

We specifically evaluate generative verifiers, which frame the verification task as a text-generation problem. These models predict "yes" or "no" tokens, with the final score derived from the normalized token probabilities. For each problem instance, we generate $K=32$ candidate rollouts and select the single highest-scoring trajectory. Performance is measured by Best@K, defined as the fraction of instances successfully resolved under this top-1 selection from the $K$ candidates.

\input{tables/verifier_results}

The results, summarized in Table \ref{tab:verifier_results}, assess several prominent open-source verifiers, including OpenHands-32B-Verifier \citep{pan2024training}, R2EGym-Verifier \citep{jain2025r2e}, and SWE-Lego-Verifier \citep{tao2026swe}. We observed significant performance gains across the \swehero suite under the TTS paradigm. Specifically, the resolution rates for the 7B and 14B models increased by 7.9\% and 5.2\%, respectively, while the 32B model saw a robust gain of 4.5\%. These results underscore the efficacy of parallel scaling in extracting higher reasoning performance from open-source agents.

\begin{figure*}[ht!]
\centering
  \begin{minipage}{\textwidth}
    \centering
    \small\textbf{Performance of SWE-Hero-7B under TTS setup.} \\ [0.5ex]
    \includegraphics[width=0.99\linewidth]{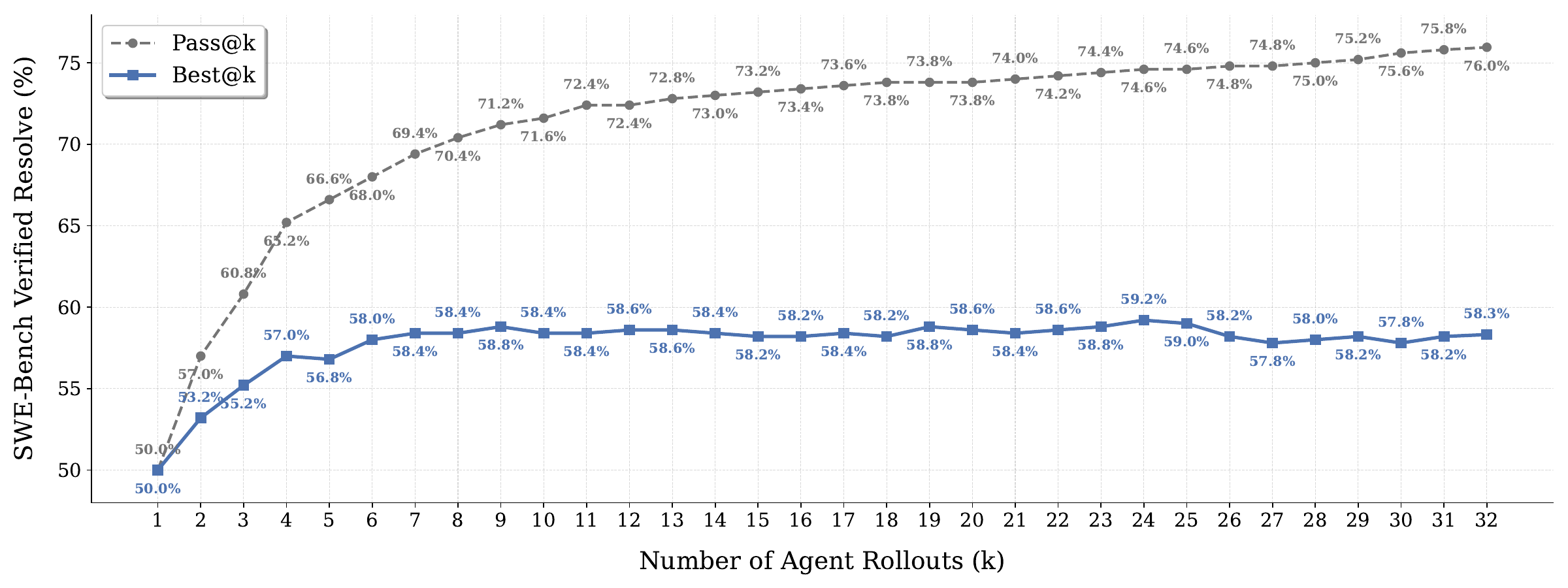}
  \end{minipage}

  \vspace{2em} 

  \begin{minipage}{\textwidth}
    \centering
    \small\textbf{Performance of SWE-Hero-14B under TTS setup.} \\ [0.5ex]
    \includegraphics[width=0.99\linewidth]{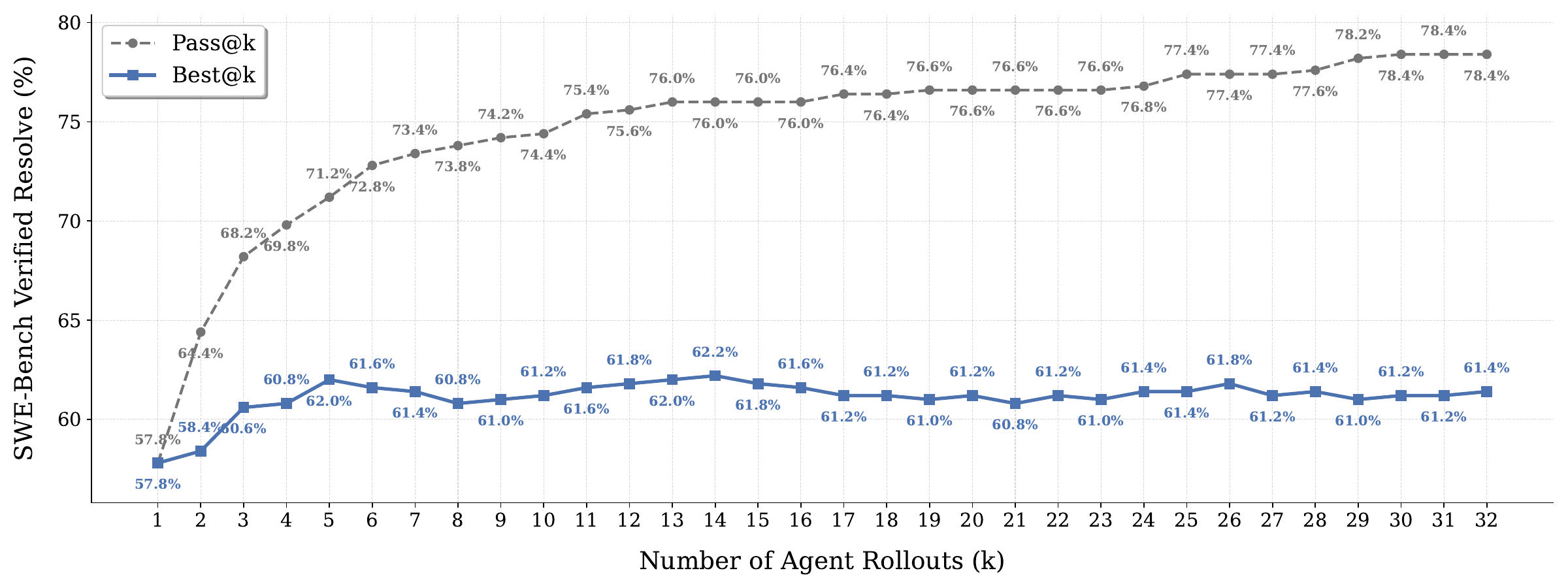}
  \end{minipage}

  \vspace{2em}

  \begin{minipage}{\textwidth}
    \centering
    \small\textbf{Performance of SWE-Hero-32B under TTS setup.} \\ [0.5ex]
    \includegraphics[width=0.99\linewidth]{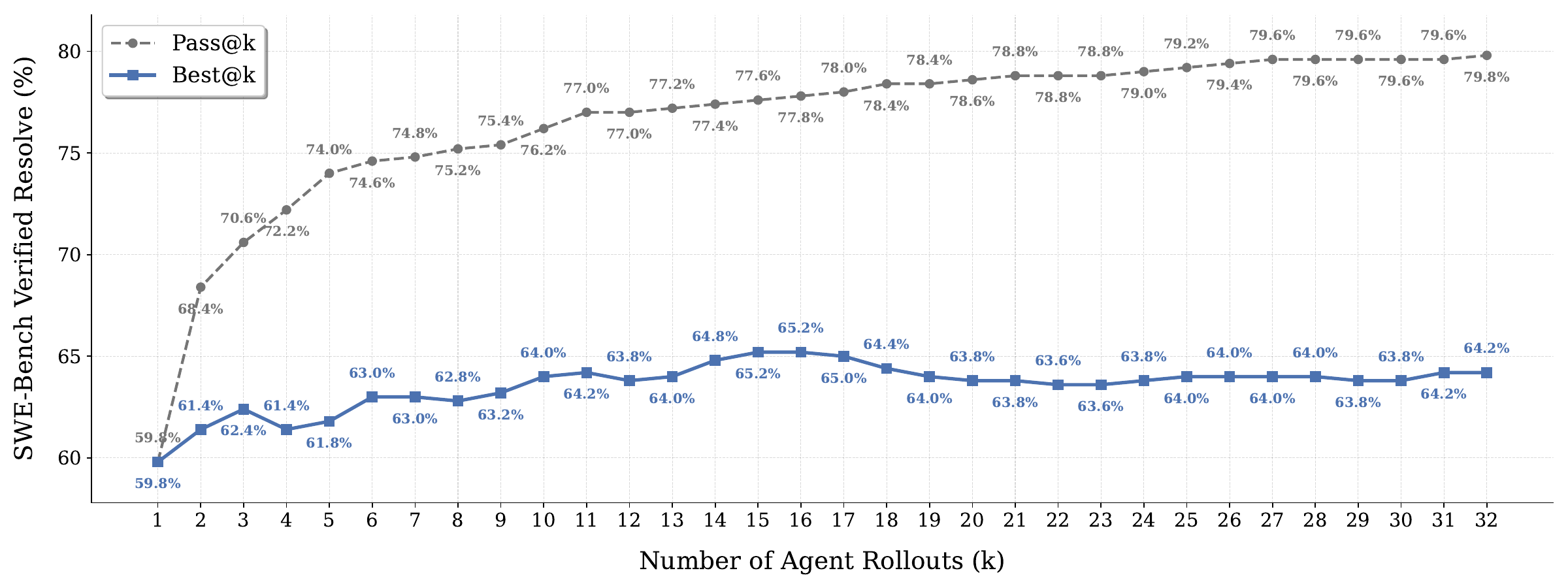}
  \end{minipage}

  \caption{
    Increasing inference-time compute improves performance on SWE-Bench Verified using open-source SWE-Lego-Verifier-8B \citep{tao2026swe}.
  }
  \label{plot:verifier_results}
\end{figure*}

However, a substantial performance overhead remains. As illustrated in Fig. \ref{plot:verifier_results}, a significant gap persists between Best@K (the verifier's selection) and Pass@K (the theoretical upper bound). Across the \swehero suite, this margin reaches as much as 15\% at $K=16$ and continues to widen as $K$ increases. This divergence indicates that current open-source verifiers still lack the discriminative precision required to bridge the gap toward the Pass@K ceiling, highlighting a critical bottleneck in the development of  autonomous software agents.

\subsection{Accuracy-to-Efficiency Trade-off: Token Consumption vs. Resolution Rate}

\begin{figure*}[ht!]
\centering
\includegraphics[width=0.99\textwidth]{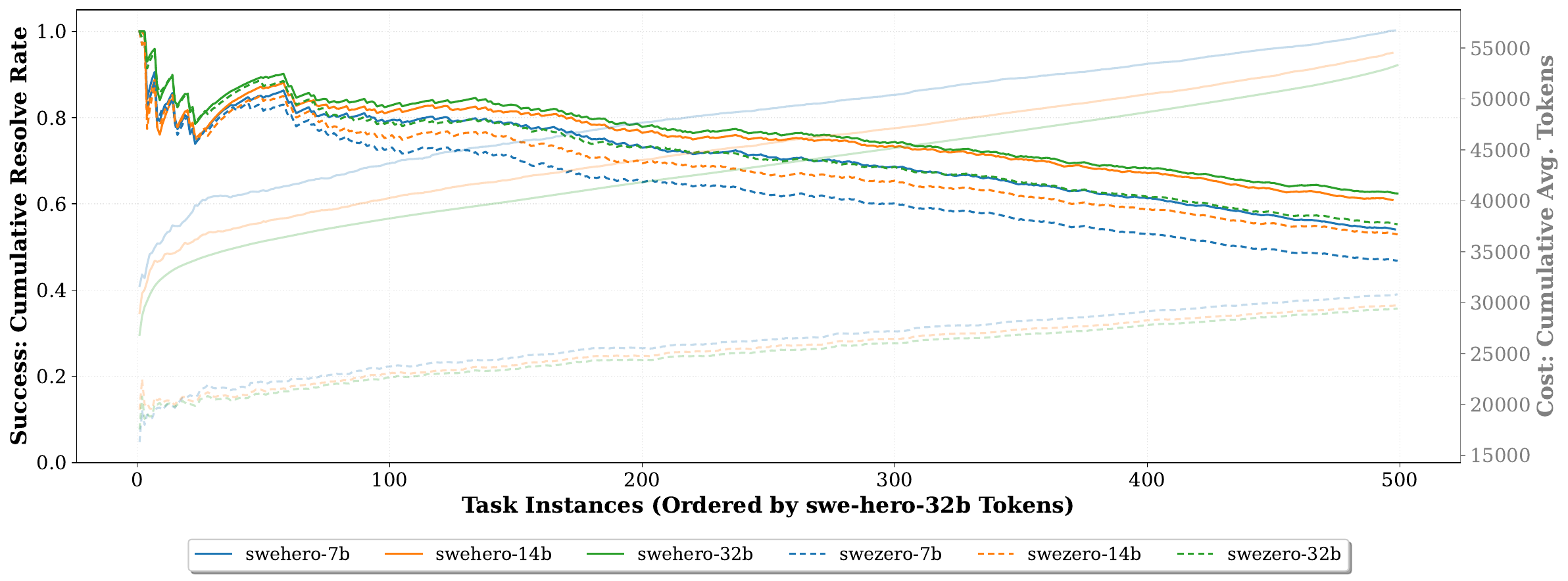}
\caption{
Cumulative performance and resource efficiency on SWE-bench Verified. Tasks are ordered by \swehero-32b token-count (complexity proxy). The primary axis (thick lines) shows the natural decline in resolve rate as complexity grows, while the secondary axis (faint lines) highlights the significantly lower token-costs of \swezero models.
}
\label{plot:tasks_vs_tokens}
\end{figure*}

We examine the efficiency-performance trade-off between computational cost—measured by total token consumption—and resolution accuracy. As shown in Figure \ref{plot:tasks_vs_tokens}, \swezero operates with extreme token efficiency by bypassing the verbose diagnostic logs and iterative test outputs inherent in execution-based environments. This makes it an ideal candidate for high-throughput applications where minimizing cost-per-resolution is the primary constraint.
In contrast, \swehero exhibits a higher token-per-task density. The agent's commitment to generating internal test suites and processing rich execution feedback naturally increases context usage and total generated tokens. However, this increased "cognitive spend" correlates directly with superior performance on complex issues. While \swehero consumes more tokens per trajectory, the investment is justified by its ability to resolve high-difficulty tasks that remain inaccessible to execution-free models.

\input{tables/swe_multilingual}

\input{tables/main_appendix}

\subsection{\swezero Trajectory Detection}

We present the \swezero trajectory detection code snippet in Fig. \ref{fig:bash_parser}.

\input{figures/exec_detection}

%% file: tables/verifier_results.tex
\begin{table}[ht]
\centering
\renewcommand{\arraystretch}{1.2} 
\resizebox{1.0\linewidth}{!} {%
\begin{tabular}{l|l|c|c|c}
\hline
\textbf{Model} & \textbf{Verifier} & \textbf{Pass@1 (avg of 32)} & \textbf{Pass@32} & \textbf{Best@32} \\ \hline

\multirow{4}{*}{SWE-Hero-7B} & OpenHands-32B-Verifier & \multirow{4}{*}{50.0} & \multirow{4}{*}{76.0} & 52.9 \\ \cdashline{2-2} \cdashline{5-5} 
 & R2EGym-Verifier &  &  & 55.9 \\ \cdashline{2-2} \cdashline{5-5} 
 & SWE-Lego-Verifier-8B &  &  & 57.9 \\ \cdashline{2-2} \cdashline{5-5} 
 & SWE-Lego-Verifier-30B-A3B &  &  & 57.9 \\ \hline

\multirow{4}{*}{SWE-Hero-14B} & OpenHands-32B-Verifier & \multirow{4}{*}{57.4} & \multirow{4}{*}{78.4} & 62.0 \\ \cdashline{2-2} \cdashline{5-5} 
 & R2EGym-Verifier &  &  & 62.6 \\ \cdashline{2-2} \cdashline{5-5} 
 & SWE-Lego-Verifier-8B &  &  & 60.6 \\ \cdashline{2-2} \cdashline{5-5} 
 & SWE-Lego-Verifier-30B-A3B &  &  & 61.0 \\ \hline

\multirow{4}{*}{SWE-Hero-32B} & OpenHands-32B-Verifier & \multirow{4}{*}{60.1} & \multirow{4}{*}{79.8} & 63.8 \\ \cdashline{2-2} \cdashline{5-5} 
 & R2EGym-Verifier &  &  & 63.8 \\ \cdashline{2-2} \cdashline{5-5} 
 & SWE-Lego-Verifier-8B &  &  & 64.6 \\ \cdashline{2-2} \cdashline{5-5} 
 & SWE-Lego-Verifier-30B-A3B &  &  & 64.0 \\ \hline

\end{tabular}
}
\caption{Evaluation results of SWE-Hero LLMs under Test Time Scaling (TTS) setting using different open-sourced generative verifiers.}
\label{tab:verifier_results}
\end{table}

%% file: tables/swe_multilingual.tex
\begin{table}[ht!]
\centering
\renewcommand{\arraystretch}{1.0} 
\begin{tabular}{llccc}
\toprule
\textbf{Model} & \textbf{Scaffold} & \textbf{Training} & \textbf{Execution} & \textbf{Resolve Rate (\%)} \\ 

\midrule
\multicolumn{5}{c}{\textbf{Proprietary Models}} \\
\hline
GPT-5.2 & mini-SWE-agent & - & $\checkmark$ & 66.3 \\
Claude 4.5 Opus & mini-SWE-agent & - & $\checkmark$ & 70.7 \\ 
Gemini 3 Flash & mini-SWE-agent & - & $\checkmark$ & 72.7 \\ 

\midrule
\multicolumn{5}{c}{\textbf{Open-Source Foundation Models}} \\
\hline
DeepSeek-V3.2 & mini-SWE-agent & - & $\checkmark$ & 62.3 \\
Qwen3-Coder-Next & SWE-agent & - & $\checkmark$ & 62.8 \\
Kimi K2.5 & mini-SWE-agent & - & $\checkmark$ & 67.3 \\
Minimax-M2.5 & mini-SWE-agent & - & $\checkmark$ & 68.3 \\
Qwen3.5-397B-A17B & Internal & - & $\checkmark$ & 69.3 \\
GLM-5 & mini-SWE-agent & - & $\checkmark$ & 69.7 \\

\midrule
\multicolumn{5}{c}{\textbf{Our Models}} \\ 
\hline
\multirow{2}{*}{SWE-Hero-7B} & \multirow{2}{*}{MOpenHands} & \multirow{2}{*}{SFT} & $\times$ & 27.9 \\
 &  &  & $\checkmark$ & 29.2 \\
\cdashline{1-5}[0.5pt/2pt]
\multirow{2}{*}{SWE-Hero-14B} & \multirow{2}{*}{MOpenHands} & \multirow{2}{*}{SFT} & $\times$ & 38.4 \\
 &  &  & $\checkmark$ & 37.5 \\
\cdashline{1-5}[0.5pt/2pt]
\multirow{2}{*}{SWE-Hero-32B} & \multirow{2}{*}{MOpenHands} & \multirow{2}{*}{SFT} & $\times$ & 42.2 \\
 &  &  & $\checkmark$ & 44.1 \\
\bottomrule
\end{tabular}
\caption{Performance comparison on SWE-bench Multilingual, comparing execution-free ($\times$) and execution-backed ($\checkmark$) settings.}
\label{tab:swe-bench-multi-results-full}
\end{table}

%% file: tables/main_appendix.tex
\begin{table}[ht!]
\centering
\renewcommand{\arraystretch}{1.2} 
\resizebox{1.0\linewidth}{!} {%
\begin{tabular}{llcc}
\toprule
\textbf{Model} & \textbf{Scaffold} & \textbf{Training} & \textbf{Resolve Rate (\%)} \\ 

\midrule
\multicolumn{4}{c}{\textbf{Proprietary Models}} \\
\hline
GPT-5.2 \citep{openai2025gpt52} & mini-SWE-agent & - & 72.8 \\
Gemini 3 Flash \citep{google2025gemini3flash} & mini-SWE-agent & - & 75.8 \\ 
Claude 4.5 Opus \citep{anthropic2025opus45} & mini-SWE-agent & - & 76.8 \\ 

\midrule
\multicolumn{4}{c}{\textbf{Open-Source Foundation Models}} \\
\hline
DeepSeek-V3.2 \citep{anthropic2025} & mini-SWE-agent & - & 70.0 \\
Qwen3-Coder-Next \citep{cao2026qwen3} & SWE-agent & - & 70.6 \\
Kimi K2.5 \citep{kimiteam2026kimik25visualagentic} & mini-SWE-agen & - & 70.8 \\
GLM-5 \citep{glm5team2026glm5} & mini-SWE-agent & - & 72.8 \\
Minimax-M2.5 \citep{minimax2026m25} & mini-SWE-agent & - & 75.8 \\
Qwen3.5-397B-A17B \citep{qwen3.5} & Internal & - & 76.4 \\

\midrule
\multicolumn{4}{c}{\textbf{Open-Source Models}} \\ 
\hline

\rowcolor{gray!10} \textit{Parameters $\approx$ 7B} & & & \\
Qwen2.5-Coder-Instruct-7B \citep{hui2024qwen2} & MOpenHands & - & 1.0 \\
Qwen3-8B \citep{yang2025qwen3} & OpenHands & - & 7.6 \\
SWE-Gym-7B \citep{pan2024training} & OpenHands & SFT & 10.6 \\
SWE-agent-LM-7B \citep{yang2025swe} & SWE-agent & SFT & 15.2 \\
Lingma-SWE-GPT-7B \citep{ma2025swe} & SWESynInfer & SFT & 18.2 \\
R2E-Gym-7B \citep{jain2025r2e} & R2E-Gym & SFT & 19.0 \\
SWE-Mirror-LM-7B \citep{wang2025swe} & MOpenHands & SFT & 22.8 \\
SWE-Dev-7B \citep{wang2025swe} & OpenHands & RL & 23.4 \\
SERA-8B \citep{shen2026sera} & SWE-agent & SFT & 31.7 \\
Klear-Agent-8B-SFT \citep{kwaiklear2025mini} & Mini-SWE-agent-plus & SFT & 39.0 \\ 
SWE-Lego-Qwen3-8B & OpenHands & SFT & 42.2$^\dagger$ \\
\rowcolor{cyan!5} SWE-Zero-7B (Ours) & OpenHands & SFT & {\bf 46.8}$^\dagger$ \\
\rowcolor{cyan!5} SWE-Hero-7B (Ours) & OpenHands & SFT & {\bf 52.7}$^\dagger$ \\
\hline

\rowcolor{gray!10} \textit{Parameters $=$ 14B} & & & \\
Qwen2.5-Coder-Instruct-14B \citep{hui2024qwen2} & OpenHands & - & 4.0 \\
SWE-Gym-14B \citep{pan2024training}  & OpenHands & SFT & 16.4 \\
Qwen3-14B \citep{yang2025qwen3} & OpenHands & - & 17.3 \\
R2E-Gym-14B \citep{jain2025r2e} & R2E-Gym & SFT & 26.8 \\
FrogMini‑14B \citep{sonwane2025bugpilot} & R2E‑Gym & SFT+RL & 45.3 \\
\rowcolor{cyan!5} SWE-Zero-14B (Ours) & OpenHands & SFT & {\bf 54.5}$^\dagger$ \\
\rowcolor{cyan!5} SWE-Hero-14B (Ours) & OpenHands & SFT & {\bf 60.8}$^\dagger$ \\
\hline

\rowcolor{gray!10} \textit{Parameters $=$ 32B} & & & \\
Qwen2.5-Coder-Instruct-32B \citep{hui2024qwen2} & MOpenHands & - & 6.2 \\
SWE-Gym-32B \citep{pan2024training}  & OpenHands & SFT & 20.6 \\
Qwen3-32B \citep{yang2025qwen3} & OpenHands & - & 23.2 \\
R2E-Gym-32B \citep{jain2025r2e} & R2E-Gym & SFT & 34.4 \\
SWE-Dev-32B \citep{wang2025swe} & OpenHands & RL & 36.6 \\
Skywork-SWE-32B \citep{zeng2025skywork} & OpenHands & SFT & 38.0 \\
SWE-agent-LM-32B \citep{yang2025swe} & SWE-agent & SFT & 40.2 \\
DeepSWE-32B-Preview \citep{luo2025deepswe} & OpenHands & RL & 42.2 \\
SWE-Mirror-LM-32B \citep{wang2025swe} & MOpenHands & SFT & 52.2 \\
SWE-Lego-Qwen3-32B \citep{tao2026swe} & OpenHands & SFT & 52.6$^\dagger$ \\
CWM-32B \citep{copet2025cwm} & Agentless & CPT + SFT + RL & 53.9 \\ 
SERA-32B \citep{shen2026sera} & SWE-agent & SFT & 54.2 \\
FrogBoss‑32B \citep{sonwane2025bugpilot} & R2E‑Gym & SFT+RL & 54.6 \\
daVinci-Dev-32B \citep{zeng2026davinci} & SWE-Agent & SFT & 56.1 \\
\rowcolor{cyan!5} SWE-Zero-32B (Ours) & OpenHands & SFT & {\bf 57.5}$^\dagger$ \\
SWE-Master-32B \citep{song2026swe} & R2E-Gym & SFT & 57.8 \\
SWE-Swiss-32B \citep{he2025sweswiss} & Agentless & SFT+RL & 58.0 \\
SWE-Master-32B-RL \citep{song2026swe} & R2E-Gym & SFT+RL & 61.4 \\
\rowcolor{cyan!5} SWE-Hero-32B (Ours) & OpenHands & SFT & {\bf 62.2}$^\dagger$ \\
OpenSWE-32B \citep{fu2026davincienvopensweenvironment} & SWE-Agent & SFT & 62.4 \\
\bottomrule
\end{tabular}
}
\caption{Performance comparison on SWE-bench Verified. Results marked with $^\dagger$ exclude \emph{Git hacking}; for all other scores, the use of \emph{Git hacking} is unspecified.}
\label{tab:swe-bench-results-appendix}
\end{table}

%% file: figures/exec_detection.tex
\begin{figure*}[!ht]
\centering
\begin{adjustbox}{valign=t,minipage=0.98\textwidth}
\begin{tabular}{l}
\lstset{
    style=CustomPy,
    basicstyle=\footnotesize\ttfamily, 
    upquote=true,
    columns=fullflexible,
    keepspaces=true,
    breaklines=true,
}
\begin{lstlisting}
import bashlex

WHITELIST = [
    "cd", "grep", "head", "find", "rm", "git", "ls", "tail", "echo", "cat", "xargs", "pwd", "mkdir", "which", "timeout", "sed", "wc", "mv", "chmod", "export", "cp", "true", "sort", "awk", "od", "printf", "xxd", "touch", "diff", "curl", "hexdump", "tr", "file", "sudo", "uniq", "basename", "cut", "sha256sum", "man", "tar", "wget"
]

class CommandNameCollector(bashlex.ast.nodevisitor):
    """Traverses the AST of a bash command to extract command names."""

    def __init__(self):
        super().__init__()
        self.names = []

    def visitcommand(self, n, parts):
        """Extracts names from command nodes and handles wrapped commands."""
        for node in parts:
            if node.kind == "word":
                self.names.append(node.word)
                break

        for i in range(len(parts)):
            if parts[i].kind == "word":
                if parts[i].word in ["sudo", "xargs", "-exec"] and i + 1 < len(parts):
                    if parts[i + 1].kind == "word":
                        self.names.append(parts[i + 1].word)
                elif parts[i].word == "timeout" and i + 2 < len(parts):
                    if parts[i + 2].kind == "word":
                        self.names.append(parts[i + 2].word)

def get_command_names(cmd: str):
    """Parses a bash string and returns a list of all detected command names."""
    if not cmd.strip():
        return []

    try:
        trees = bashlex.parse(cmd)
    except Exception:
        return None

    command_names = []
    for tree in trees:
        collector = CommandNameCollector()
        collector.visit(tree)
        command_names.extend(collector.names)
    return command_names

def is_prohibited_command(cmd: str) -> bool:
    """Returns True if the command contains names not found in the WHITELIST."""
    command_names = get_command_names(cmd)
    if command_names is None:
        return True
    return any(name not in WHITELIST for name in command_names)

\end{lstlisting}
\end{tabular}
\end{adjustbox}

\caption{
Detection of execution in the \texttt{execute\_bash} command. This Python implementation leverages \texttt{bashlex} to generate an Abstract Syntax Tree (AST), ensuring all invoked commands are validated against a predefined \texttt{WHITELIST}.
}
\label{fig:bash_parser}
\end{figure*}